\documentclass[a4paper,fleqn]{cas-dc}

\usepackage[numbers]{natbib}
\usepackage{subfigure}
\usepackage{tikz}
\usepackage{url}
\usepackage{xcolor}

\begin{document}
\let\WriteBookmarks\relax
\def\floatpagepagefraction{1}
\def\textpagefraction{.001}

\shorttitle{Leveraging Self-Sovereign Identity, Blockchain, and Zero-Knowledge Proof to Build a Privacy-Preserving Vaccination Pass}

\shortauthors{M Barros et~al.}

\title [mode = title]{Leveraging Self-Sovereign Identity, Blockchain, and Zero-Knowledge Proof to Build a Privacy-Preserving Vaccination Pass}                      

\author[1]{Maurício Barros}[orcid=0000-0001-6825-892X]
\cormark[1]
\ead{mauricio.v.barros@gmail.com}
\address[1]{Programa de Pós-Graduação em Ciência da Computação, Departamento de Informática e Estatística, Universidade Federal de Santa Catarina, 88040-370, Florianópolis, Brazil}

\author[1,2]{Frederico Schardong}[orcid=0000-0003-4492-163X]
\fnmark[1]
\ead{frede.sc@gmail.com}
\ead{frederico.schardong@rolante.ifrs.edu.br}
\address[2]{Instituto Federal de Educação, Ciência e Tecnologia do Rio Grande do Sul - Campus Rolante, 95690-000, Rolante, Brazil}

\author[1]{Ricardo Felipe Custódio.}[orcid=0000-0001-9611-5694]
\ead{ricardo.custodio@ufsc.br}

\cortext[cor1]{Corresponding author}

\fntext[fn1]{This study was supported by the Federal Institute of Education, Science and Technology of Rio Grande do Sul (IFRS).}

\begin{abstract}
The current humanitarian health crisis popularized the debate on data privacy. At the same time, several cities, states, and even countries put the mandatory presentation of health pass to access services into practice. In this article, we explore the concepts of self-sovereign identity, blockchain, and zero-knowledge proofs to propose a solution to the problem of presenting proof of vaccination. This solution allows users to prove that they are vaccinated for different pathogens without revealing their identity. The architecture is loosely coupled, allowing components to be exchanged, which we discuss when we present the implementation of a working prototype.
\end{abstract}




\maketitle

\section{Introduction}

The COVID-19 pandemic surprised the world, generating an enormous health crisis with profound social and economic implications. Reopening the economy becomes a latent issue as many countries are vaccinating considerable portions of their populations~\cite{owidcoronavirus}.

The transition period between the prohibition of access to services and the complete and unrestricted reopening of the economy has been marked by the use of proof of vaccination for the SARS-CoV-2 virus, commonly called COVID passports, or COVID pass~\cite{9558786}. Countries, states, and cities are now demanding proof of vaccination, as is the case in the city of São Paulo~\cite{pass} and the state of New York~\cite{nys}. Such vaccination certificates help economic recovery, allowing vaccinated people to access services that invariably require close contact with others.

However, the use of vaccination certificates poses privacy-related challenges. For example, you need to ensure that whoever presents a vaccination certificate is who they say they are. It is not trivial to solve this problem without de-anonymizing the certificate submitter. Another challenge is to allow a vaccinated person, in addition to proving their vaccination anonymously, not to reveal which laboratory manufactured their vaccine or even the exact date of vaccination.

Vaccination certificates can be made in various ways, from a physical version of paper or plastic card to electronic. While the physical version is more vulnerable to fraud and doesn't solve the challenges listed above, it's cheap and readily available to everyone. The electronic version, which presupposes access to technology, can offer greater privacy, security, and scalability if properly designed and developed, possibly successfully tackling the challenges described above. One promising way to attack such challenges electronically is to use Self-Sovereign Identity (SSI).

SSI is a user-centric paradigm of digital identity~\cite{allen,schardong2021selfsovereign}. In an SSI-based ecosystem, the user is central to identity management, having control and cryptographic guarantees over shared personal data. The SSI concept is especially relevant in sensitive personal data, such as health data. In addition to GDPR~\cite{gdpr2016}, health data generally have specific legislation, such as HIPAA~\cite{hipaa}, which establishes a legal framework requiring special care regarding the storage, handling, and sharing of health information.

We tackled the challenge of creating a digital vaccination pass that respects users' privacy in this work. It allows their proof of vaccination to be verified while reducing the exposure of personal and health data. Our SSI solution to this challenge maintains user anonymity through blockchain and Zero-Knowledge Proof (ZKP).

This article is organized as follows. In Section~\ref{sec:id}, we present the concepts of Digital Identity and Self-Sovereign Identity. Section~\ref{sec:con} introduces concepts required to follow this work, namely Verifiable Credentials (VCs), ZKP, Decentralized Identifiers (DIDs), and Blockchain. We present our proposal and detail our implementation in Sections~\ref{sec:proposal} and \ref{sec:solution}, respectively. In Section~\ref{sec:related}, related works are discussed, and in Section~\ref{sec:conclusion}, we close the article presenting our conclusions.

\section{Digital Identity}\label{sec:id}

\begin{figure*}[tp]
    \centering  
    \subfigure[Centralized]  
    {  
        \begin{tikzpicture}[x=0.43pt,y=0.43pt,yscale=-1,xscale=1]
            \draw   (20,1715) .. controls (20,1695.67) and (35.67,1680) .. (55,1680) .. controls (74.33,1680) and (90,1695.67) .. (90,1715) .. controls (90,1734.33) and (74.33,1750) .. (55,1750) .. controls (35.67,1750) and (20,1734.33) .. (20,1715) -- cycle ;
            \draw   (180,1715) .. controls (180,1695.67) and (195.67,1680) .. (215,1680) .. controls (234.33,1680) and (250,1695.67) .. (250,1715) .. controls (250,1734.33) and (234.33,1750) .. (215,1750) .. controls (195.67,1750) and (180,1734.33) .. (180,1715) -- cycle ;
            \draw    (93,1715) -- (177,1715) ;
            \draw [shift={(180,1715)}, rotate = 540] [fill={rgb, 255:red, 0; green, 0; blue, 0 }  ][line width=0.08]  [draw opacity=0] (8.93,-4.29) -- (0,0) -- (8.93,4.29) -- cycle    ;
            \draw [shift={(90,1715)}, rotate = 360] [fill={rgb, 255:red, 0; green, 0; blue, 0 }  ][line width=0.08]  [draw opacity=0] (8.93,-4.29) -- (0,0) -- (8.93,4.29) -- cycle    ;
            
            \draw (55,1715) node   [align=left] {User};
            \draw (215,1702) node   [align=left] {SP};
            \draw (215,1728) node   [align=left] {IdP};
        
        \end{tikzpicture}
        \label{fig:iam:centralized}
    }  
    \subfigure[Outsourced IdP]  
    {  
        \begin{tikzpicture}[x=0.43pt,y=0.43pt,yscale=-1,xscale=1]
            \draw   (20,1945) .. controls (20,1925.67) and (35.67,1910) .. (55,1910) .. controls (74.33,1910) and (90,1925.67) .. (90,1945) .. controls (90,1964.33) and (74.33,1980) .. (55,1980) .. controls (35.67,1980) and (20,1964.33) .. (20,1945) -- cycle ;
            \draw   (180,1945) .. controls (180,1925.67) and (195.67,1910) .. (215,1910) .. controls (234.33,1910) and (250,1925.67) .. (250,1945) .. controls (250,1964.33) and (234.33,1980) .. (215,1980) .. controls (195.67,1980) and (180,1964.33) .. (180,1945) -- cycle ;
            \draw   (100,1825) .. controls (100,1805.67) and (115.67,1790) .. (135,1790) .. controls (154.33,1790) and (170,1805.67) .. (170,1825) .. controls (170,1844.33) and (154.33,1860) .. (135,1860) .. controls (115.67,1860) and (100,1844.33) .. (100,1825) -- cycle ;
            \draw    (110.32,1855.08) -- (73.28,1909.72) ;
            \draw [shift={(71.6,1912.2)}, rotate = 304.13] [fill={rgb, 255:red, 0; green, 0; blue, 0 }  ][line width=0.08]  [draw opacity=0] (8.93,-4.29) -- (0,0) -- (8.93,4.29) -- cycle    ;
            \draw [shift={(112,1852.6)}, rotate = 124.13] [fill={rgb, 255:red, 0; green, 0; blue, 0 }  ][line width=0.08]  [draw opacity=0] (8.93,-4.29) -- (0,0) -- (8.93,4.29) -- cycle    ;
            \draw    (159.66,1855.1) -- (196.34,1910.5) ;
            \draw [shift={(198,1913)}, rotate = 236.49] [fill={rgb, 255:red, 0; green, 0; blue, 0 }  ][line width=0.08]  [draw opacity=0] (8.93,-4.29) -- (0,0) -- (8.93,4.29) -- cycle    ;
            \draw [shift={(158,1852.6)}, rotate = 56.49] [fill={rgb, 255:red, 0; green, 0; blue, 0 }  ][line width=0.08]  [draw opacity=0] (8.93,-4.29) -- (0,0) -- (8.93,4.29) -- cycle    ;
            \draw    (177,1945) -- (93,1945) ;
            \draw [shift={(90,1945)}, rotate = 360] [fill={rgb, 255:red, 0; green, 0; blue, 0 }  ][line width=0.08]  [draw opacity=0] (8.93,-4.29) -- (0,0) -- (8.93,4.29) -- cycle    ;
            \draw [shift={(180,1945)}, rotate = 180] [fill={rgb, 255:red, 0; green, 0; blue, 0 }  ][line width=0.08]  [draw opacity=0] (8.93,-4.29) -- (0,0) -- (8.93,4.29) -- cycle    ;
            \draw  [dash pattern={on 0.84pt off 2.51pt}]  (168.07,1847.49) -- (210,1910) ;
            \draw [shift={(166.4,1845)}, rotate = 56.15] [fill={rgb, 255:red, 0; green, 0; blue, 0 }  ][line width=0.08]  [draw opacity=0] (8.93,-4.29) -- (0,0) -- (8.93,4.29) -- cycle    ;

            \draw (55,1945) node   [align=left] {User};
            \draw (215,1945) node   [align=left] {SP};
            \draw (135,1825) node   [align=left] {IdP};
        \end{tikzpicture}
        \label{fig:iam:outsourced}
    }
    \subfigure[User-centric]{
        \begin{tikzpicture}[x=0.43pt,y=0.43pt,yscale=-1,xscale=1]
            \draw   (20,2085) .. controls (20,2065.67) and (35.67,2050) .. (55,2050) .. controls (74.33,2050) and (90,2065.67) .. (90,2085) .. controls (90,2104.33) and (74.33,2120) .. (55,2120) .. controls (35.67,2120) and (20,2104.33) .. (20,2085) -- cycle ;
            \draw   (180,2085) .. controls (180,2065.67) and (195.67,2050) .. (215,2050) .. controls (234.33,2050) and (250,2065.67) .. (250,2085) .. controls (250,2104.33) and (234.33,2120) .. (215,2120) .. controls (195.67,2120) and (180,2104.33) .. (180,2085) -- cycle ;
            \draw    (93,2085) -- (177,2085) ;
            \draw [shift={(180,2085)}, rotate = 180] [fill={rgb, 255:red, 0; green, 0; blue, 0 }  ][line width=0.08]  [draw opacity=0] (8.93,-4.29) -- (0,0) -- (8.93,4.29) -- cycle    ;
            \draw [shift={(90,2085)}, rotate = 0] [fill={rgb, 255:red, 0; green, 0; blue, 0 }  ][line width=0.08]  [draw opacity=0] (8.93,-4.29) -- (0,0) -- (8.93,4.29) -- cycle    ;
            \draw  [dash pattern={on 0.84pt off 2.51pt}]  (93,2070) -- (180,2070) ;
            \draw [shift={(90,2070)}, rotate = 0] [fill={rgb, 255:red, 0; green, 0; blue, 0 }  ][line width=0.08]  [draw opacity=0] (8.93,-4.29) -- (0,0) -- (8.93,4.29) -- cycle    ;
        
            \draw (55,2098) node   [align=left] {User};
            \draw (215,2085) node   [align=left] {SP};
            \draw (55,2072) node   [align=left] {IdP};
        \end{tikzpicture}
        \label{fig:iam:usercentric}
    }
    \subfigure[SSI]{
        \begin{tikzpicture}[x=0.43pt,y=0.43pt,yscale=-1,xscale=1]
            \draw   (170,2325) .. controls (170,2305.67) and (185.67,2290) .. (205,2290) .. controls (224.33,2290) and (240,2305.67) .. (240,2325) .. controls (240,2344.33) and (224.33,2360) .. (205,2360) .. controls (185.67,2360) and (170,2344.33) .. (170,2325) -- cycle ;
            \draw   (90,2205) .. controls (90,2185.67) and (105.67,2170) .. (125,2170) .. controls (144.33,2170) and (160,2185.67) .. (160,2205) .. controls (160,2224.33) and (144.33,2240) .. (125,2240) .. controls (105.67,2240) and (90,2224.33) .. (90,2205) -- cycle ;
            \draw    (100.32,2235.08) -- (63.28,2289.72) ;
            \draw [shift={(61.6,2292.2)}, rotate = 304.13] [fill={rgb, 255:red, 0; green, 0; blue, 0 }  ][line width=0.08]  [draw opacity=0] (8.93,-4.29) -- (0,0) -- (8.93,4.29) -- cycle    ;
            \draw [shift={(102,2232.6)}, rotate = 124.13] [fill={rgb, 255:red, 0; green, 0; blue, 0 }  ][line width=0.08]  [draw opacity=0] (8.93,-4.29) -- (0,0) -- (8.93,4.29) -- cycle    ;
            \draw    (167,2325) -- (83,2325) ;
            \draw [shift={(80,2325)}, rotate = 360] [fill={rgb, 255:red, 0; green, 0; blue, 0 }  ][line width=0.08]  [draw opacity=0] (8.93,-4.29) -- (0,0) -- (8.93,4.29) -- cycle    ;
            \draw [shift={(170,2325)}, rotate = 180] [fill={rgb, 255:red, 0; green, 0; blue, 0 }  ][line width=0.08]  [draw opacity=0] (8.93,-4.29) -- (0,0) -- (8.93,4.29) -- cycle    ;
            \draw   (10,2325) .. controls (10,2305.67) and (25.67,2290) .. (45,2290) .. controls (64.33,2290) and (80,2305.67) .. (80,2325) .. controls (80,2344.33) and (64.33,2360) .. (45,2360) .. controls (25.67,2360) and (10,2344.33) .. (10,2325) -- cycle ;
            \draw  [dash pattern={on 0.84pt off 2.51pt}]  (149.66,2235.1) -- (188,2293) ;
            \draw [shift={(148,2232.6)}, rotate = 56.49] [fill={rgb, 255:red, 0; green, 0; blue, 0 }  ][line width=0.08]  [draw opacity=0] (8.93,-4.29) -- (0,0) -- (8.93,4.29) -- cycle    ;
            \draw  [dash pattern={on 0.84pt off 2.51pt}]  (83,2310) -- (170,2310) ;
            \draw [shift={(80,2310)}, rotate = 0] [fill={rgb, 255:red, 0; green, 0; blue, 0 }  ][line width=0.08]  [draw opacity=0] (8.93,-4.29) -- (0,0) -- (8.93,4.29) -- cycle    ;
            
            \draw (205,2325) node   [align=left] {SP};
            \draw (125,2205) node   [align=left] {IdP};
            \draw (45,2338) node   [align=left] {User};
            \draw (45,2312) node   [align=left] {IdP};
        \end{tikzpicture}
        \label{fig:iam:ssi}
    }
    \caption{The four IAM models. Constant lines represent interactions, and dashed lines mean trust~\cite{schardong2021selfsovereign}}
    \label{fig:iam}
\end{figure*}
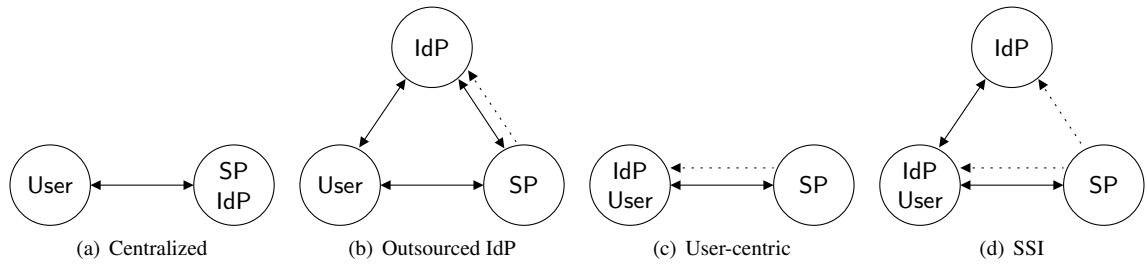

ISO 24760-1, which addresses security and privacy for identity management, defines digital identity as ``a set of attributes related to an entity''~\cite{iso2019}. In~\cite{4385303}, digital identity is defined as the representation of an entity linked to a specific context. Linking digital identity to specific contexts is a commonly accepted idea in the literature~\cite{8776589,schardong2021selfsovereign,josang2005user}.

Identity management also called Identity and Access Management (IAM), is a set of policies, processes, and technologies related to the administration of digital identities~\cite{iso2019}. An identity management system defines how entities are identified, authenticated, and authorized to access restricted access services~\cite{8776589}. Digital identity and IAMs can be modeled in different ways. In this section, we describe the three traditional models of digital identity~\cite{allen}, pointing out their problems to, finally, explain SSI and how this fourth model attacks the issues of the previous models.

\subsection{Isolated Identity}

The isolated model was the first identity model. It is the simplest model~\cite{8776589}, as depicted in Figure~\ref{fig:iam:centralized}. In this paradigm, only the user and the service provider (SP) exist, and the SP also operates as an Identity Provider (IdP). Each web service that requires authentication and authorization must implement its own IAM.

One of the consequences of this model is that users have many digital identities spread across different online services. Furthermore, each SP has to bear the costs of implementing and maintaining an IAM, which involves protecting against possible attacks, vulnerabilities, and taking care of the demands imposed by GDPR.

\subsection{Outsourced IdP}

The natural evolution of the previous model is the separation of IAM functionalities into a specific service for this purpose, giving rise to the identity provider as a service in itself. In this way, the IdP positions itself as an intermediary between the user and the SP, as shown in Figure~\ref{fig:iam:outsourced}. In this model, the SP outsources identity management to the IdP.

This model tackles the problems of (i) usability, \textit{i.e.} users having many accounts, and; (ii) responsibility of SPs, \textit{i.e.} SPs having to design, implement and care for sensitive user information. In this model, users do not need to have several distinct identities, and SPs do not need to implement their own IAM framework.

However, this model generated unexpected consequences, such as an immense concentration of data in the hands of a few IdPs, \textit{e.g.} Facebook, Google, and Twitter. These companies have become substantial data silos, trapping people in an oligarchy of few IDPs without portability among themselves~\cite{schardong2021selfsovereign}.

By concentrating high amounts of personal data, these silos have become centers of attention, decoys for attacks. In addition, issues such as data leaks, security, data property, and privacy raise concerns. The data belongs to users, but they do not own and control them and are unaware of all services that consume them.

\subsection{User-Centric}

One of the answers to the problem of users having to hand over their data to an IdP was given by the user-centric identity model~\cite{josang2005user}, illustrated in Figure~\ref{fig:iam:usercentric}. The proposal is for the user to store access credentials issued by SPs in a personal authentication device, such as a smartcard or smartphone. The user authenticates to the personal authentication device using a PIN code or another method, and the authentication device, in turn, authenticates with the SP.

Although this model advances privacy, allowing the user to control their access credentials, it does not address the management of user attributes or the incorporation of attributes guaranteed by third parties. The latter issue is of great relevance, as many SPs will only trust the value of attributes if they are issued by the IdPs they trust. For example, a car rental company will not accept a driver's license if the user issues it.

\subsection{Self-Sovereign Identity}

Although a precise definition of SSI is still under debate, as discussed in \cite{schardong2021selfsovereign}, \cite{Muhle2018} and \cite{8776589}, the SSI literature solidifies the idea that the user should own and manage their data. In other words, both attributes and credentials, which can be self-signed or signed by third parties, must be controlled only by users, who present their data to SPs whenever they desire. This is depicted in Figure~\ref{fig:iam:ssi}. Therefore, users can choose to have few or many digital personas, and no silos of personal data are created.

In 2016, Christopher Allen proposed ten guiding principles for SSI~\cite{allen}, and although it is a blog post, it is treated as a whitepaper in the area~\cite{schardong2021selfsovereign}. The ten principles are as follows. (i) Existence; the user must exist independent of any service provider/identity provider. (ii) Control; the user must control their identities. (iii) Access; the user must always have access to their data. (iv) Transparency; all systems and algorithms must be transparent to the user. (v) Persistence; identities must be persisted. (vi) Portability; the services and information about the user's identity must be transportable. (vii) Interoperability; identities should have as wide a range of use as possible across different systems. (viii) Consent: The user must always consent to the usage of his identity data. (ix) Minimization; when necessary to show some information, always show as little as possible to accomplish the task. (x) Protection; users' rights must be protected.

While it is possible to implement SSI without Blockchain~\cite{vanbokkem2019selfsovereign} technology, using it brings significant benefits. For example, storing credential revocation schemes on the ledger brings high availability and an immutable record of revoked data. Another advantage is to remove third parties when establishing trust by allowing entities to be part of the network and prove they are who they say they are in novel ways. For instance, a customer physically accessing a bank agency scans a QR code posted on the wall and connects to the bank, thus not requiring the bank to have an x509 certificate within a PKI hierarchy nor an IANA-controlled DNS domain. Finally, using blockchain to implement SSI fosters an ontologically coherent ecosystem, as it publishes credential metadata that can be reused and augmented by all participants.

\section{Terminology and Concepts}\label{sec:con}

This section introduces the fundamental concepts needed to understand this work. These are the decentralized identifiers, verifiable credentials standards, the concept of Zero-Knowledge Proofs (ZKP), and blockchain.

\subsection{Decentralized Identifiers}

SSI aims to give users control over their data. Part of this effort consists of ensuring that communications in peer-to-peer relations do not involve third parties. However, trust in today's communications over the internet is rooted in third-party authorities: chains of certificates tied to root certificate authorities (CAs) chosen \textit{a priori} by browsers and operating systems. The decentralized identifiers (DIDs) standard is being developed at W3C to mitigate the participation of authorities in communications and relationships in SSI~\cite{did2021}.

DIDs are designed to operate independently of centralized registries, identity providers, and CAs. Every DID is linked to a unique DID document, a JSON structure that can be stored on-chain or off-chain, that describes an entity, specifying public keys, service endpoints, etc. For instance, an entity can be a person, a group, a relationship between entities, an organization, or an internet of things object~\cite{did2021}.

The format of a DID identifier is a textual string composed of three parts separated by colons, where the first part is always \texttt{DID}; the second is a method identifier; the third is an entity identifier in this method. A DID method is a technique that describes the creation and management of method-specific identifiers and how to obtain their respective DID documents. For example, the \texttt{did:indy} method describes identifiers on Hyperledger Indy blockchains, while the \texttt{did:key} method describes identifiers generated from cryptographic keys. An example of DID address of this method is \texttt{did:key:z6MkpTHR8VN sBxYAAWHut2Geadds9dVWtWAnuB}.

\subsection{Verifiable Credentials}

Working with user attributes is a crucial part of any system that operates with digital identities. Attributes are commonly organized into structures called credentials, which are also used for identification and authentication. A credential contains a set of one or more claims made by an issuer about an entity~\cite{Muhle2018}. More formally, a credential is a set of claims and not attributes because claims can be false or represent an incomplete perception of the whole. In contrast, attributes are properties, \textit{i.e.} absolute truths, of entities.

Verifiable Credential (VC) is a set of claims and metadata that can cryptographically prove the identity of the issuer through a digital signature, providing integrity and authenticity~\cite{vc2019}. VCs have metadata that describes the issuer, expiration date, public key, and revocation information. Regarding the latter, the cryptographic accumulator is often employed to create private-preserving revocation registries. The cryptographic accumulator is an algorithm that combines a set of values into one short accumulator, such that evidence is produced that the accumulator incorporated a given value without revealing it.~\cite{fazio2002cryptographic}. 

Part of the W3C VC standard defines the verifiable presentation (VP) of claims~\cite{vc2019}. VP is the process of revealing one or more claims to a verifier. The verifier may or may not trust the claims presented but must always be able to verify integrity and authenticity. VPs can also reveal the result of operators on claims, for instance, that someone's birth date was at least 18 years ago. This technique is called Zero-Knowledge Proof (ZKP).

\subsection{Zero-Knowledge Proof}

Zero-Knowledge Proof (ZKP) is a cryptographic protocol where one proves to someone else that they know a value, but without revealing any other information other than the fact that they know it~\cite{schnorr1989efficient}. Formally, ZKP is an interactive proofing system $(P, V)$ for proving a language membership statement over a language $L=\{0,1\}^*$ with two actors, the prover $P$ and the verifier $V$. To prove that an instance $x \in L$, $P$ and $V$ must share $x$, denoted by $(P, V)(x)$. Then, $P$ and $V$ exchange a sequence of messages so that at the end of the interactions, the result is $(P, V)(x) \in \{accept, reject\}$, representing the acceptance or not of $V$ for the statement of $P$ that $x \in L$~\cite{6206859}.

ZKP schemes must have two properties~\cite{goldreich1994definitions}: (i) completeness, an interactive proof is complete if the participants are honest, \textit{i.e.} follow the protocol correctly, and the protocol succeeds with overwhelming probability; and (ii) soundness, an interactive proof is sound if a dishonest prover can only mislead $V$ with negligible probability. ZKPs are not proofs in the mathematical sense as there is a small probability of convincing $V$ of a false statement. Thus, multiple protocol instances are used to reduce the likelihood to a negligible probability of convincing $V$ of a false statement.

In the context of VCs, ZKPs are used to build VPs that convince the verifier about something concerning the claims it contains without revealing them. For instance, a driver's license was issued to me and is valid, or my credit score is higher than a given threshold. To reduce the communication between the prover and the verifier, Non-Interactive Zero-Knowledge Proof (NIZKP)~\cite{micali2000computationally} is used. NIZKP enables proofs to be built and sent via an out-of-band method, such as a QR code. This technique uses Fiat-Shamir paradigm~\cite{fiat1986prove} to remove interactions from the protocol.

\subsection{Blockchain}

Most SSI systems~\cite{Muhle2018} use distributed ledger technology to: (i) decentralize storage and, consequently, reduce authoritarian control over data; and (ii) have guaranteed immutability of the information stored in the ledger. 

Those features are possible because of the properties of blockchain data structures such as append-only hash lists and Merkle trees~\cite{merkle1987digital}. In a blockchain, blocks of data contain, together with its data, the hash of the previous block. Thus, to modify the data of a single block within a sequence of linked blocks (\textit{i.e.} a chain), one would need to change all subsequent blocks.

Having an append-only structure guarantees immutability on a local level. Nonetheless, this structure must be distributed in a network, and all participant nodes must agree on the correct values of the blocks. This agreement is achieved through a consensus algorithm. Most solutions for this problem are attempts on solving the Byzantine generals problem~\cite{lamport2019byzantine}. The most popular solution to this problem are algorithms based on proof of work, popularised by Bitcoin~\cite{nakamoto}. 

\section{Digital Vaccination Pass}\label{sec:proposal}

The undertaking of designing a credential for COVID vaccination is complex. Such a credential must comply with different actors from different areas, both local and international. Several study groups and collaborative efforts~\cite{ghpc2021,cci2021,who2022} were formed around the world to design or standardize these vaccination certificates. One such initiative, the Good Health Pass Collaborative (GHPC)~\cite{ghpc2021} is a multi-sector initiative to create a blueprint for the interoperability of digital health pass systems~\cite{ghpcblue2021}.

In their first White Paper, the GHPC initiative defined four critical requirements that health credential systems must satisfy~\cite{ghpcwhite2021}: (i) the credential must be able to work across borders and comply with local and international legislation; (ii) the credential must have the collaboration of different areas such as health, governments, tourism, travel, etc; (iii) the credential must comply with privacy and data protection regulations and must be able to link to the credential holder; and (iv) the credential shall not add costs or other burdens to users. This work aims to comply with these requirements, adopting open-source tools and open standards as a way to make this technology available.

SSI introduces essential ideas and principles regarding people's privacy. The ten principles are especially relevant in the contemporary context of the COVID-19 pandemic, which has accelerated the digitization of society and popularized the debate over the privacy of health data~\cite{9558786}. Based on the ten SSI principles, the concepts discussed above, and the requirements laid down by the GHPC, it is possible to tackle the challenge of allowing a person to prove their vaccination status while preserving their privacy. In other words, to prove whether or not one is vaccinated for a disease without revealing when or where they were vaccinated or even hide the laboratory that produced the vaccine.

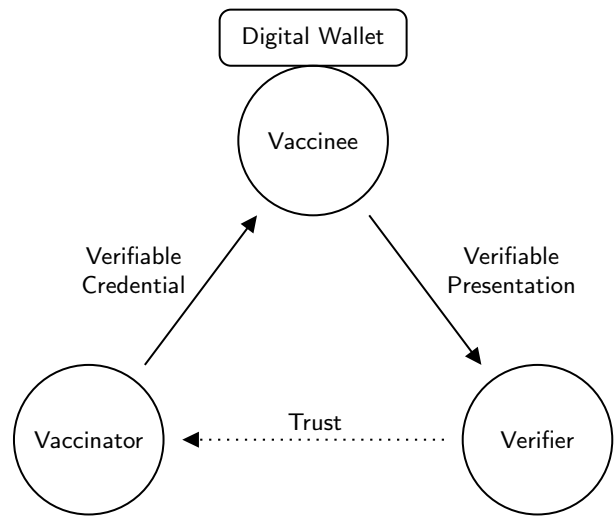
\begin{figure}[h]
  \centering
  \tikzset{every picture/.style={line width=0.75pt}} 

\begin{tikzpicture}[x=0.7pt,y=0.7pt,yscale=-1,xscale=1]

\draw   (50,260) .. controls (50,237.91) and (67.91,220) .. (90,220) .. controls (112.09,220) and (130,237.91) .. (130,260) .. controls (130,282.09) and (112.09,300) .. (90,300) .. controls (67.91,300) and (50,282.09) .. (50,260) -- cycle ;
\draw   (290,260) .. controls (290,237.91) and (307.91,220) .. (330,220) .. controls (352.09,220) and (370,237.91) .. (370,260) .. controls (370,282.09) and (352.09,300) .. (330,300) .. controls (307.91,300) and (290,282.09) .. (290,260) -- cycle ;
\draw   (170,100) .. controls (170,77.91) and (187.91,60) .. (210,60) .. controls (232.09,60) and (250,77.91) .. (250,100) .. controls (250,122.09) and (232.09,140) .. (210,140) .. controls (187.91,140) and (170,122.09) .. (170,100) -- cycle ;
\draw   (160,36) .. controls (160,32.69) and (162.69,30) .. (166,30) -- (254,30) .. controls (257.31,30) and (260,32.69) .. (260,36) -- (260,54) .. controls (260,57.31) and (257.31,60) .. (254,60) -- (166,60) .. controls (162.69,60) and (160,57.31) .. (160,54) -- cycle ;
\draw    (120,220) -- (178.2,142.4) ;
\draw [shift={(180,140)}, rotate = 486.87] [fill={rgb, 255:red, 0; green, 0; blue, 0 }  ][line width=0.08]  [draw opacity=0] (8.93,-4.29) -- (0,0) -- (8.93,4.29) -- cycle    ;
\draw    (240,140) -- (298.2,217.6) ;
\draw [shift={(300,220)}, rotate = 233.13] [fill={rgb, 255:red, 0; green, 0; blue, 0 }  ][line width=0.08]  [draw opacity=0] (8.93,-4.29) -- (0,0) -- (8.93,4.29) -- cycle    ;
\draw  [dash pattern={on 0.84pt off 2.51pt}]  (280,260) -- (143,260) ;
\draw [shift={(140,260)}, rotate = 360] [fill={rgb, 255:red, 0; green, 0; blue, 0 }  ][line width=0.08]  [draw opacity=0] (8.93,-4.29) -- (0,0) -- (8.93,4.29) -- cycle    ;

\draw (210,45) node   [align=left] {Digital Wallet};
\draw (210,100) node   [align=left] {Vaccinee};
\draw (90,260) node   [align=left] {Vaccinator};
\draw (330,260) node   [align=left] {Verifier};
\draw (114,169) node   [align=left] {\begin{minipage}[lt]{49.22pt}\setlength\topsep{0pt}
\begin{center}
Verifiable\\Credential
\end{center}

\end{minipage}};
\draw (316,169) node   [align=left] {\begin{minipage}[lt]{60pt}\setlength\topsep{0pt}
\begin{center}
Verifiable\\Presentation
\end{center}

\end{minipage}};
\draw (211,250.5) node   [align=left] {Trust};

\end{tikzpicture}  
  \caption{The three roles and their relations in our solution.}
  \label{fig:triade}
\end{figure}

Our proposal for a proof of vaccination uses the concepts discussed above. It starts with a public or private entity previously authorized to administer vaccines, namely a vaccinator. It is registered on the blockchain to issue VCs for vaccinated people, \textit{i.e.} the vaccinees. Since any entity on the blockchain can issue VCs, it is necessary to use VC issuing authorization schemes as proposed in~\cite{lauinger2021poa} to ensure that only VCs issued by authorized entities are legitimately recognized. Figure~\ref{fig:triade} illustrates the three entities involved in our solution.

The VC states the vaccinee's full name and date of birth, the laboratory that produced the administered vaccine, the applied dose (first, second, third, or more if applicable), and the date it was administered. Our VC format does not include country-specific identifiers, such as the SSN of USA or South Corea's Resident's Registration Number, for two reasons: (i) our solution is country independent; and (ii) there are countries with large numbers of unregistered citizens~\cite{id4d}.

The vaccinee stores the VC in a digital wallet. This digital wallet can be a smartwatch, a smartphone app that uses a secure hardware element, or another Personal Assistant Device (PAD). Sending the VC from the vaccinator to the vaccinee takes place immediately after the act of vaccination through the most convenient and secure way for the user. For instance, if the PAD has Near Field Communication (NFC) and Bluetooth, NFC is favored because of the low range of the protocol. QR code scanning is also available. Regardless of the communication technology used to connect the digital wallet with the issuer, DIDs represent the entities involved and identify the relationship between these two entities.

Note that the credential is held by the user in their wallet, causing no personal data to be stored on the blockchain, and therefore the system has no GDPR compliance obligations. Personal data are in possession of their owner, and their consent is required for sharing them. This arrangement makes it difficult for issuing organizations to track user data activities. In addition, in the hands of the owners, it is an excellent way to transport this data across borders without the need for substantial centralized data centers.

When needed, the vaccinee can use her PAD's digital wallet to produce a VP, proving that she is vaccinated to a verifier. The user can customize the VP to have more or less personal data. In this way, the amount of exposed data can be adjusted to suit the context better. For instance, suppose the vaccinee does not wish to reveal the laboratory that produced her vaccine. In this case, she can create a VP using ZKP to prove that the value of the laboratory field on her VC is one of the values in a list of laboratories accepted by the verifier.

Alternatively, the verifier might define the VP format. For instance, suppose that to access a social event such as a concert or play, the verifier (\textit{i.e.} the entity responsible for the event) needs to ensure that participants have taken at least one dose of the COVID-19 vaccine. The verifier defines a VP request with a specific format in this case. They send it to the customers, who, through their digital wallets, accept or not to produce a VP following the requested format. An airline, however, may be obliged to demand complete vaccination of its passengers 14 days before the flight. In this case, the airline constructs a VP request for this context, and passengers agree or disagree to produce a VP when boarding the plane.

Finally, the verifier validates the VP and chooses to trust it if they trust the vaccinator who issued the VC. Verifiers can confirm the authentication and authorization of vaccinators through the DID registered in the blockchain. The blockchain also stores the value of the cryptographic accumulator, which is used to create a revocation registry of VCs, which can be used if, for example, expired vaccine doses have been applied by mistake.



\section{Empirical Experimentation}\label{sec:solution}

The shift of focus regarding data ownership introduced by SSI allows us to propose a privacy-preserving digital vaccination pass. This section presents the tools and technologies adopted to realize our goal, an architectural overview of our system, and specific implementation details and results.

\subsection{Underlying Technologies}

We previously presented concepts, standards, and technologies, and now we discuss the tools employed in our solution and how they work. With regards to the blockchain, we use a blockchain specially created for identity management in the SSI model named Hyperledger Indy. It is an open-source project of the Hyperledger community that provides an ecosystem for SSI based on Blockchain~\cite{indy}. The Indy distributed ledger consists of two subprojects, \texttt{Indy-Plenum} and \texttt{Indy-Node}. While the former is a general-purpose blockchain that implements the consensus algorithm, the latter is a specialization of the former, where identity-specific transactions are implemented~\cite{shcherbakov2019}. It is important to note that Indy does not store personal data on the blockchain. The data saved on the ledger are DIDs of entities that issue VCs, their public keys, contact endpoints, a cryptographic accumulator value to serve as a revocation list, and VC schema metadata.

Using blockchain here has different advantages. It is possible to create an international chain of trust through decentralization, with various governments issuing their lists of trusted DIDs. It also makes it challenging to have a point of failure that would stop the system. Also, a centralized personal and health data repository cannot be created because there is no personal data on the blockchain or elsewhere.

To foster and facilitate the adoption of SSI, the Hyperledger community has also created Hyperledger Aries~\cite{aries}. It is an abstraction layer above Indy, implementing methods to produce, transfer and store VCs and VPs independent of the blockchain solution below. Through this abstraction, developers can focus on business rules and not worry about implementation details. The abstraction layer that Aries introduces happens through a software agent called the Aries agent.

The Aries agent interacts with other entities via DID Communication (DIDComm) or other communication protocols. DIDComm is a protocol that allows asynchronous communication of DIDs through encrypted messages~\cite{didcomm}. Two parts make up the Aries agent, namely the agent and the controller. The former is responsible for creating, signing, and reading transactions on the blockchain, interacting with other agents, managing secure storage, creating and presenting VPs using ZKP, and exchanging messages with the controller. The latter implements business logic, indicating how the agent should respond to events. The agent and controller communicate via REST API. The agent sends HTTP webhook calls to the controller, which in turn analyzes and responds to the agent accordingly~\cite{become}.

\subsection{System Overview}

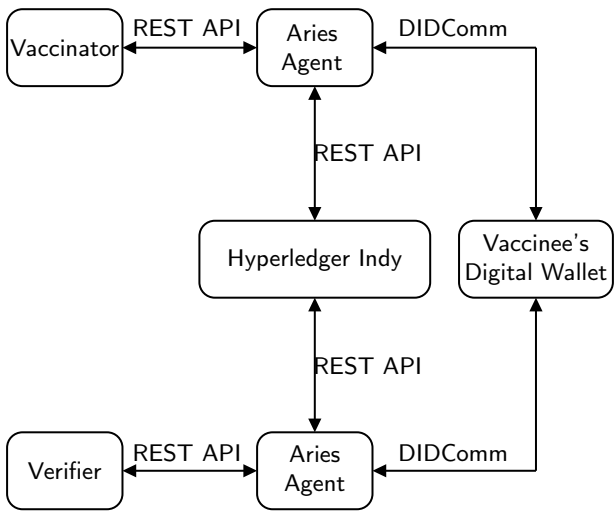
\begin{figure}[h]
  \centering
  \tikzset{every picture/.style={line width=0.75pt}} 

\begin{tikzpicture}[x=0.72pt,y=0.72pt,yscale=-1,xscale=1]

\draw   (70,28) .. controls (70,23.58) and (73.58,20) .. (78,20) -- (122,20) .. controls (126.42,20) and (130,23.58) .. (130,28) -- (130,52) .. controls (130,56.42) and (126.42,60) .. (122,60) -- (78,60) .. controls (73.58,60) and (70,56.42) .. (70,52) -- cycle ;
\draw   (200,28) .. controls (200,23.58) and (203.58,20) .. (208,20) -- (252,20) .. controls (256.42,20) and (260,23.58) .. (260,28) -- (260,52) .. controls (260,56.42) and (256.42,60) .. (252,60) -- (208,60) .. controls (203.58,60) and (200,56.42) .. (200,52) -- cycle ;
\draw    (133,40) -- (197,40) ;
\draw [shift={(200,40)}, rotate = 180] [fill={rgb, 255:red, 0; green, 0; blue, 0 }  ][line width=0.08]  [draw opacity=0] (7.14,-3.43) -- (0,0) -- (7.14,3.43) -- cycle    ;
\draw [shift={(130,40)}, rotate = 0] [fill={rgb, 255:red, 0; green, 0; blue, 0 }  ][line width=0.08]  [draw opacity=0] (7.14,-3.43) -- (0,0) -- (7.14,3.43) -- cycle    ;
\draw   (70,248) .. controls (70,243.58) and (73.58,240) .. (78,240) -- (122,240) .. controls (126.42,240) and (130,243.58) .. (130,248) -- (130,272) .. controls (130,276.42) and (126.42,280) .. (122,280) -- (78,280) .. controls (73.58,280) and (70,276.42) .. (70,272) -- cycle ;
\draw   (200,248) .. controls (200,243.58) and (203.58,240) .. (208,240) -- (252,240) .. controls (256.42,240) and (260,243.58) .. (260,248) -- (260,272) .. controls (260,276.42) and (256.42,280) .. (252,280) -- (208,280) .. controls (203.58,280) and (200,276.42) .. (200,272) -- cycle ;
\draw    (133,260) -- (197,260) ;
\draw [shift={(200,260)}, rotate = 180] [fill={rgb, 255:red, 0; green, 0; blue, 0 }  ][line width=0.08]  [draw opacity=0] (7.14,-3.43) -- (0,0) -- (7.14,3.43) -- cycle    ;
\draw [shift={(130,260)}, rotate = 0] [fill={rgb, 255:red, 0; green, 0; blue, 0 }  ][line width=0.08]  [draw opacity=0] (7.14,-3.43) -- (0,0) -- (7.14,3.43) -- cycle    ;
\draw   (170,138) .. controls (170,133.58) and (173.58,130) .. (178,130) -- (282,130) .. controls (286.42,130) and (290,133.58) .. (290,138) -- (290,162) .. controls (290,166.42) and (286.42,170) .. (282,170) -- (178,170) .. controls (173.58,170) and (170,166.42) .. (170,162) -- cycle ;
\draw    (230,63) -- (230,127) ;
\draw [shift={(230,130)}, rotate = 270] [fill={rgb, 255:red, 0; green, 0; blue, 0 }  ][line width=0.08]  [draw opacity=0] (7.14,-3.43) -- (0,0) -- (7.14,3.43) -- cycle    ;
\draw [shift={(230,60)}, rotate = 90] [fill={rgb, 255:red, 0; green, 0; blue, 0 }  ][line width=0.08]  [draw opacity=0] (7.14,-3.43) -- (0,0) -- (7.14,3.43) -- cycle    ;
\draw    (230,237) -- (230,173) ;
\draw [shift={(230,170)}, rotate = 450] [fill={rgb, 255:red, 0; green, 0; blue, 0 }  ][line width=0.08]  [draw opacity=0] (7.14,-3.43) -- (0,0) -- (7.14,3.43) -- cycle    ;
\draw [shift={(230,240)}, rotate = 270] [fill={rgb, 255:red, 0; green, 0; blue, 0 }  ][line width=0.08]  [draw opacity=0] (7.14,-3.43) -- (0,0) -- (7.14,3.43) -- cycle    ;
\draw    (263,40) -- (345,40) ;
\draw [shift={(260,40)}, rotate = 0] [fill={rgb, 255:red, 0; green, 0; blue, 0 }  ][line width=0.08]  [draw opacity=0] (7.14,-3.43) -- (0,0) -- (7.14,3.43) -- cycle    ;
\draw    (263,260) -- (345,260) ;
\draw [shift={(260,260)}, rotate = 0] [fill={rgb, 255:red, 0; green, 0; blue, 0 }  ][line width=0.08]  [draw opacity=0] (7.14,-3.43) -- (0,0) -- (7.14,3.43) -- cycle    ;
\draw    (345,40) -- (345,127) ;
\draw [shift={(345,130)}, rotate = 270] [fill={rgb, 255:red, 0; green, 0; blue, 0 }  ][line width=0.08]  [draw opacity=0] (7.14,-3.43) -- (0,0) -- (7.14,3.43) -- cycle    ;
\draw    (345,260) -- (345,173) ;
\draw [shift={(345,170)}, rotate = 450] [fill={rgb, 255:red, 0; green, 0; blue, 0 }  ][line width=0.08]  [draw opacity=0] (7.14,-3.43) -- (0,0) -- (7.14,3.43) -- cycle    ;
\draw   (305,138) .. controls (305,133.58) and (308.58,130) .. (313,130) -- (377,130) .. controls (381.42,130) and (385,133.58) .. (385,138) -- (385,162) .. controls (385,166.42) and (381.42,170) .. (377,170) -- (313,170) .. controls (308.58,170) and (305,166.42) .. (305,162) -- cycle ;

\draw (100,40) node   [align=left] {\begin{minipage}[lt]{50.16pt}\setlength\topsep{0pt}
\begin{center}
Vaccinator
\end{center}

\end{minipage}};
\draw (228.5,41) node   [align=left] {\begin{minipage}[lt]{29.38pt}\setlength\topsep{0pt}
\begin{center}
Aries\\Agent
\end{center}

\end{minipage}};
\draw (100,260) node   [align=left] {\begin{minipage}[lt]{34.47pt}\setlength\topsep{0pt}
\begin{center}
Verifier
\end{center}

\end{minipage}};
\draw (230,260) node   [align=left] {\begin{minipage}[lt]{29.38pt}\setlength\topsep{0pt}
\begin{center}
Aries\\Agent
\end{center}

\end{minipage}};
\draw (230,150) node   [align=left] {Hyperledger Indy};
\draw (163.5,29.5) node   [align=left] {REST API};
\draw (163.5,250.5) node   [align=left] {REST API};
\draw (257.5,205.5) node   [align=left] {REST API};
\draw (257.5,94.5) node   [align=left] {REST API};
\draw (302,29.5) node   [align=left] {DIDComm};
\draw (302,250.5) node   [align=left] {DIDComm};
\draw (345,150) node   [align=left] {\begin{minipage}[lt]{61.86pt}\setlength\topsep{0pt}
\begin{center}
Vaccinee's\\Digital Wallet
\end{center}

\end{minipage}};

\end{tikzpicture}  
  \caption{System architecture.}
  \label{fig:arquiProto}
\end{figure}

We have previously presented the three roles involved in the vaccine pass and their interactions. We now detail the architecture of our system and how these roles concretely interact using the technologies presented. Figure~\ref{fig:arquiProto} shows the architecture of our system.

First, it's important to note that the first time an organization instantiates our solution, some preparation is required. Specifically, an Aries agent must communicate with the Indy blockchain to: (i) record the DID of each health agency that will administer vaccines; (ii) define the format of the VCs that will be issued; and (iii) specify the revocation registration of the VCs. To ensure the chain of trust, DIDs issued to health agencies must be endorsed by the government or the responsible agency. The government then makes available a list of trusted DIDs, contributing to a global chain of trust.

After carrying out the preparations described above, each health agency authorized to administer vaccines must control an Aries agent through software, mobile app, or website. The Aries agent can run on the same device that controls it or remotely from an organization's or third-party cloud data center. When a health agency representative vaccinates someone, an HTTP call via REST API is made to the health agency's agent requesting the issuance of a VC. The vaccinee's name and date of birth are sent in the request, along with information about the applied vaccine such as manufacturing laboratory, pathogen the vaccine fights, dose number, place, and date of application. The information sent to the agent can be customized.

The Aries agent controlled by the health agency stores the private key of the entity it represents in a secure enclave and uses it to issue a VC for the received data. The newly created VC belongs to the vaccinee and must be forwarded to him. A peer-to-peer connection using DIDComm is made between the vaccinator's Aries agent and the vaccinee's digital wallet to transfer the VC. It is important to note that the blockchain only stores the value of the cryptographic accumulator, to prove the unrevoked status of the VC and nothing else.

Finally, the vaccinee can create a VP to prove something about his VC as described earlier. The secure environment of the digital wallet creates the VP and sends it via DIDComm to the Aries agent of the verifier. The verifier's agent connects to Indy to check whether the VC that originated the VP is revoked or not. All data exchanges are private and peer-to-peer, between issuer and holder and between holder and verifier.

\subsection{Prototype Implementation}

We employ the tools described above to implement a privacy-preserving solution to the problem of proving vaccination status, which is available online\footnote{URL omitted for peer-review}. With regards to the blockchain, our prototype uses Sovrin, an implementation of Hyperledger Indy that acts as a network of interoperable SSI networks~\cite{sov}. It is one of the first SSI offerings and perhaps the most studied~\cite{Kuperberg2019, Muhle2018, Lim2018}, allowing different SSI systems to share VC metadata and to interoperate, thus fostering the adoption of SSI.

We created two websites that adapt to desktops and mobile devices using \texttt{HTML} and \texttt{JavaScript} for the proof-of-concept implementation. The vaccinator operates one website to issue VCs, while the verifier uses the second website to define VP formats and present this request to vaccinees. We integrated the backend of each website with an Aries agent. We chose to use a proprietary Aries agent implementation called Verity because it is readily available and runs in the cloud. It is a product of Evernym, the company that created Indy and donated it to the Hyperledger Foundation~\cite{tobin2016inevitable}. Nonetheless, there are free and open-source Aries agents such as \texttt{ACA-Py}~\cite{acapy}.

\begin{figure}[h]
  \centering
  \includegraphics[width=.55\linewidth]{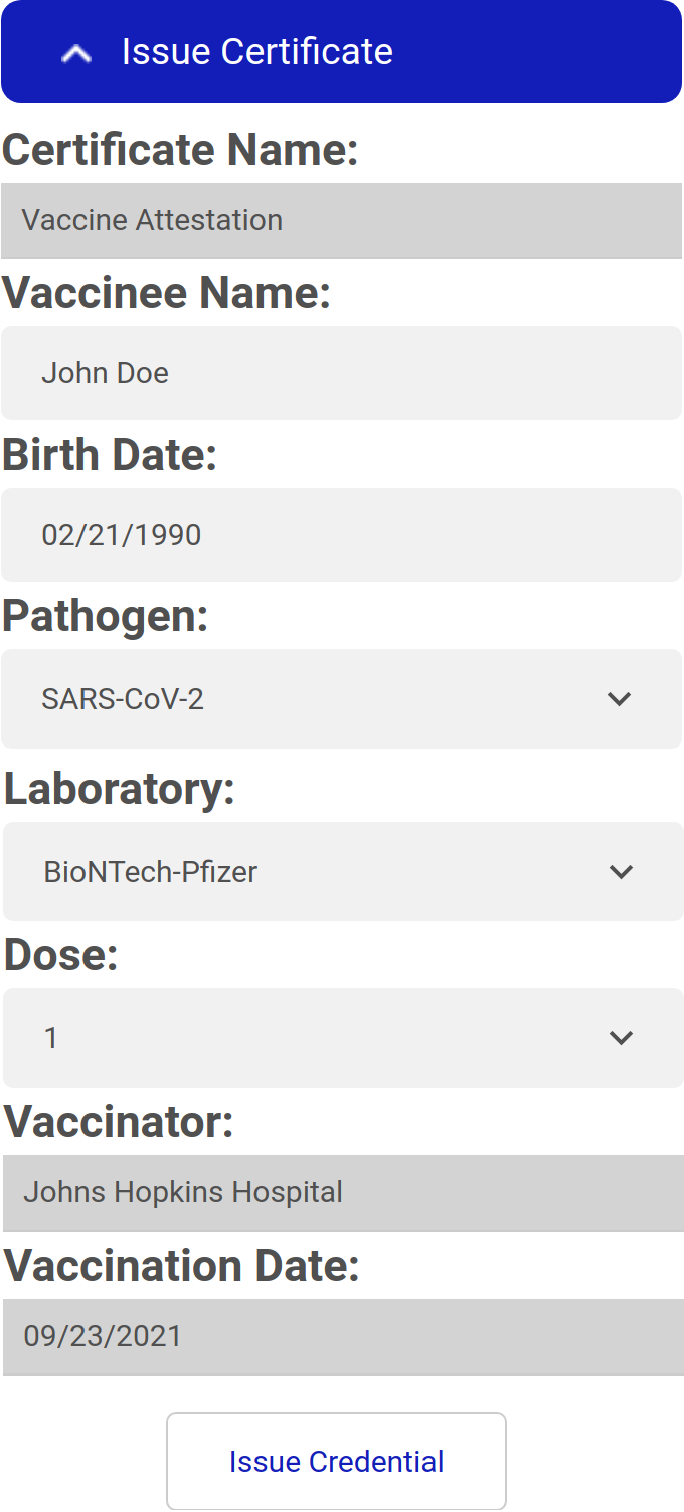}
  \caption{The website where the vaccinator fills in the vaccine and vaccinee data.}
  \label{fig:telaEmissaoVC}
\end{figure}

\begin{figure}[h]
  \centering
  \includegraphics[width=.5\linewidth]{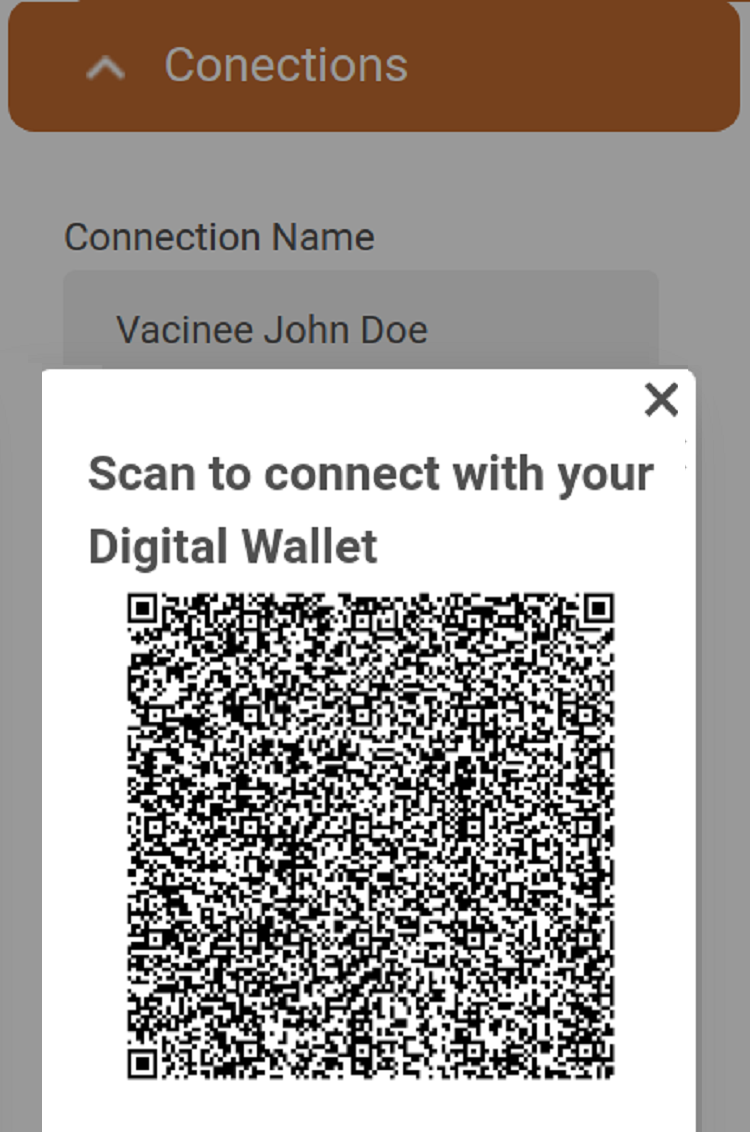}
  \caption{VC transfer via QR code. Vaccinee needs to scan the QR code with their digital wallet.}
  \label{fig:qrcode}
\end{figure}

Once a vaccine is applied, the vaccine and vaccinee data are submitted to the vaccinator's website as shown in Figure~\ref{fig:telaEmissaoVC}. In the background, the system requests its Aries agent to create a VC and transfer it to the vaccinee's digital wallet. The transference begins with the vaccinee reading a QR code that is returned by our application, as shown in Figure~\ref{fig:qrcode}.
    
    
\begin{figure}
    \centering 
    \subfigure[Digital wallet requesting confirmation to connect with the vaccinator.]{\label{fig:a}\includegraphics[width=41mm]{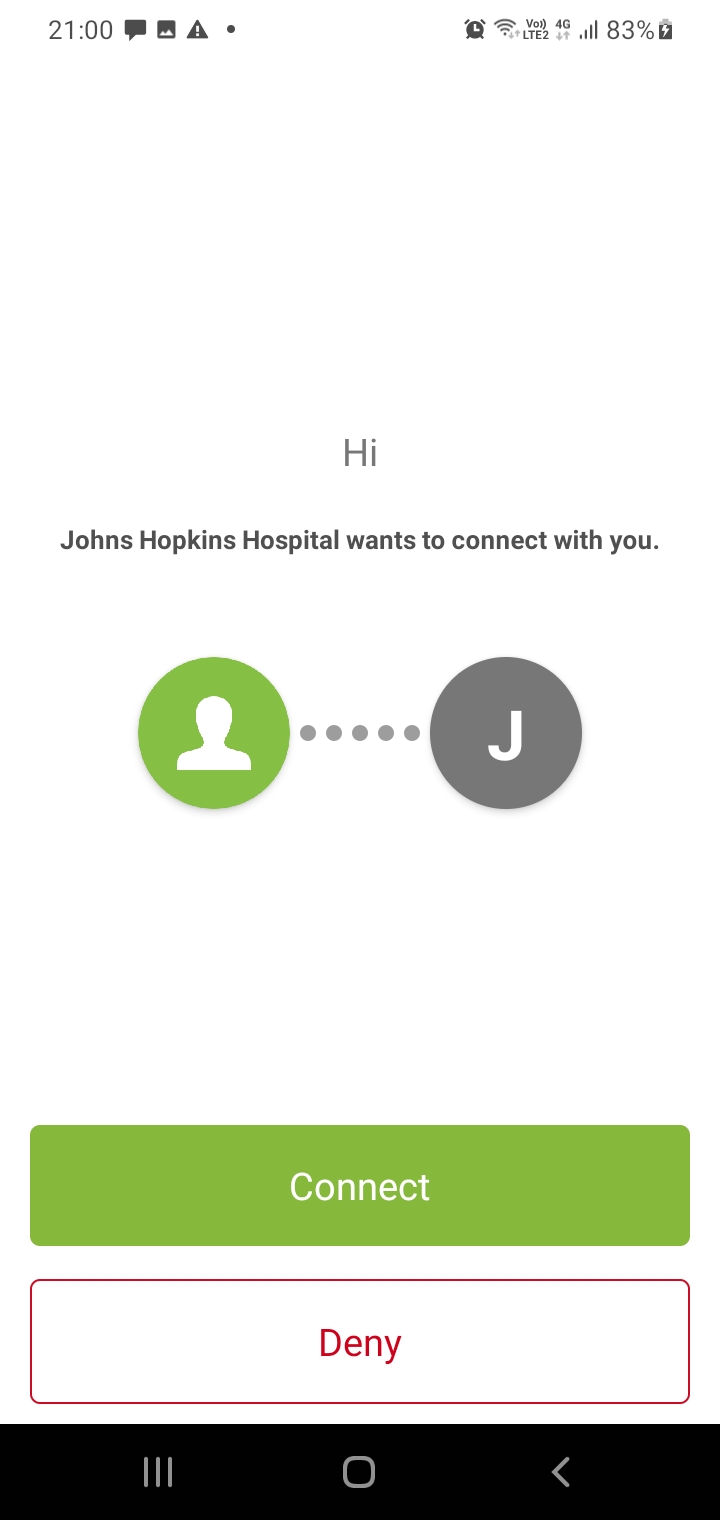}}
    \subfigure[Digital wallet requesting confirmation to receive the VC.]{\label{fig:b}\includegraphics[width=41mm]{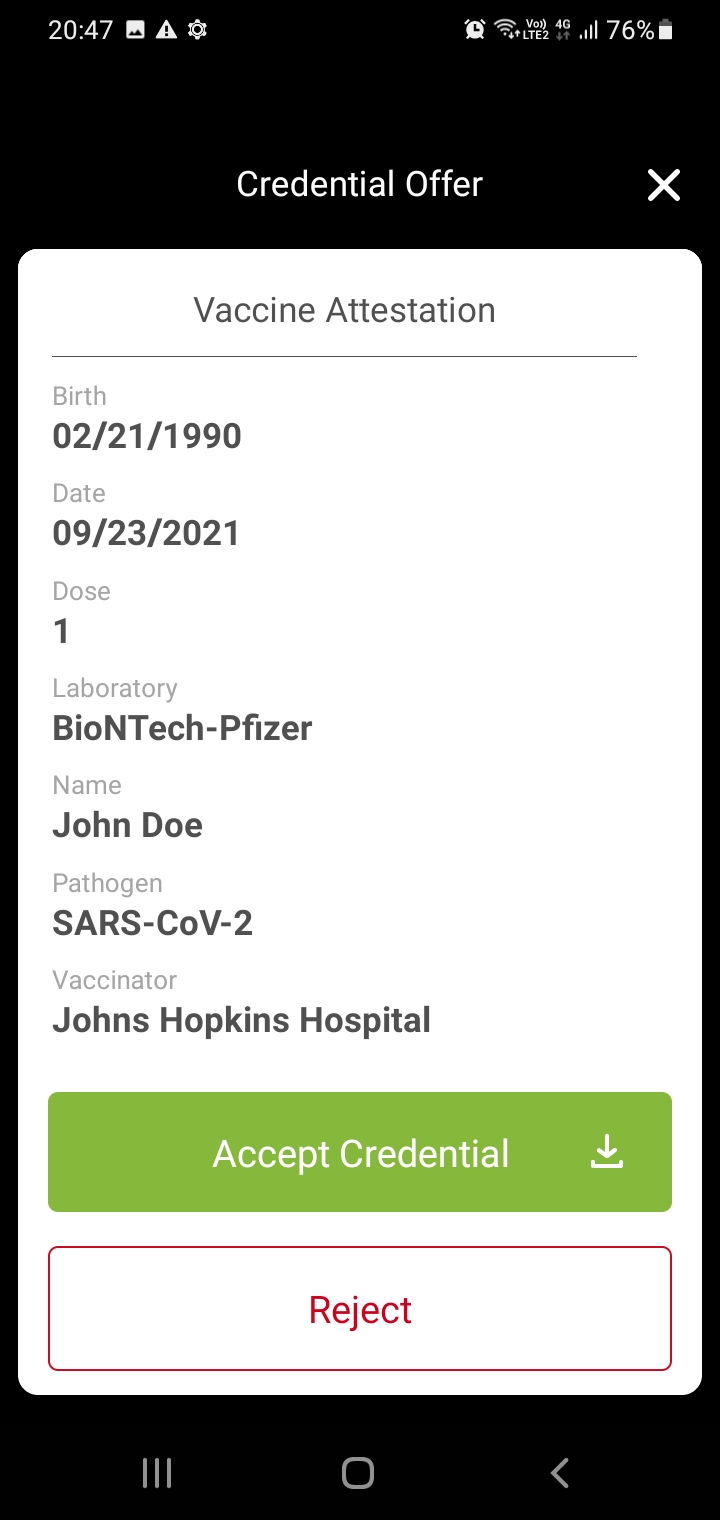}}
    \caption{Vaccinee connects to vaccinator through \texttt{DIDComm} and receives her VC.}
\end{figure}

In our implementation, we used the digital wallet \texttt{connect.me} to perform the actions of a vaccinee~\cite{connect}. There are a variety of free and paid digital wallets capable of receiving VCs and producing VPs~\cite{wallets}. Figure~\ref{fig:a} shows \texttt{connect.me} asking for user confirmation to either accept or deny connecting to the vaccinator after scanning the QR code. Although the wallet does not inform how this connection happens to the user, it is a peer-to-peer connection using \texttt{DIDComm}. Figure~\ref{fig:b} shows the wallet asking the vaccinee to accept or deny receiving their VC, which occurs immediately after confirming the incoming connection from the vaccinator. Upon acceptance, the vaccinee can use their digital wallet to produce VPs and prove their vaccination status.

Establishing a biometric link between the wallet and the vaccinee is imperative in all transactions. For instance, when the issuer connects to the vaccinee to send a VC, the issuer must mandate that the wallet is secured with a biometric factor, such as fingerprint or facial recognition. Likewise, when the verifier checks a VP, the wallet that created the VP must be secured with a biometric factor, ensuring that the wallet holder is, in fact, the owner of the VC. In the \texttt{connect.me} wallet, blocking and unblocking by biometric factors is available.

\begin{figure}[h]
  \centering
  \includegraphics[width=.55\linewidth]{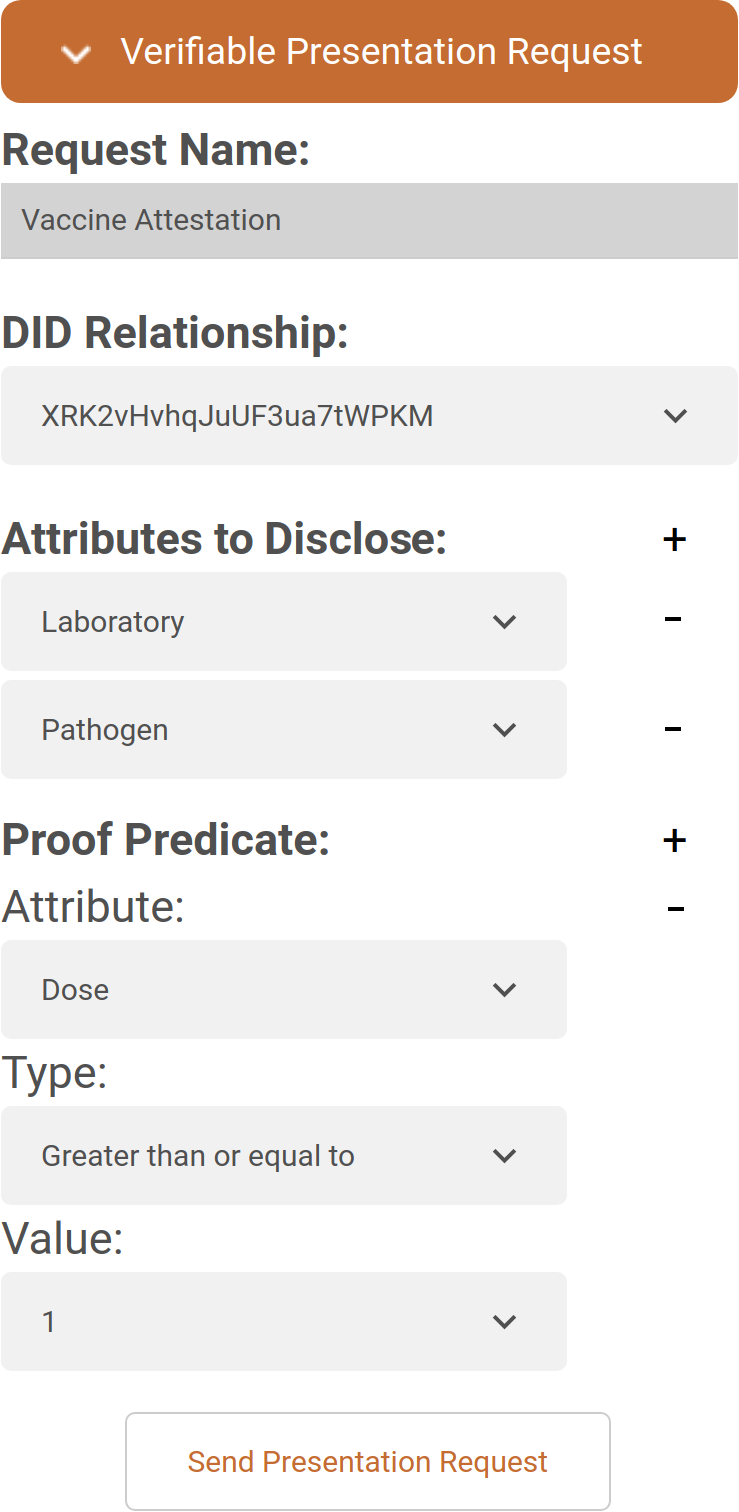}
  \caption{Request for proof of first dose.}
  \label{requisicaoProva}
\end{figure}

\begin{figure}[h]
    \centering 
    \subfigure[Digital wallet asks the vaccinee to share a VP.]{\label{fig:c}\includegraphics[width=41mm]{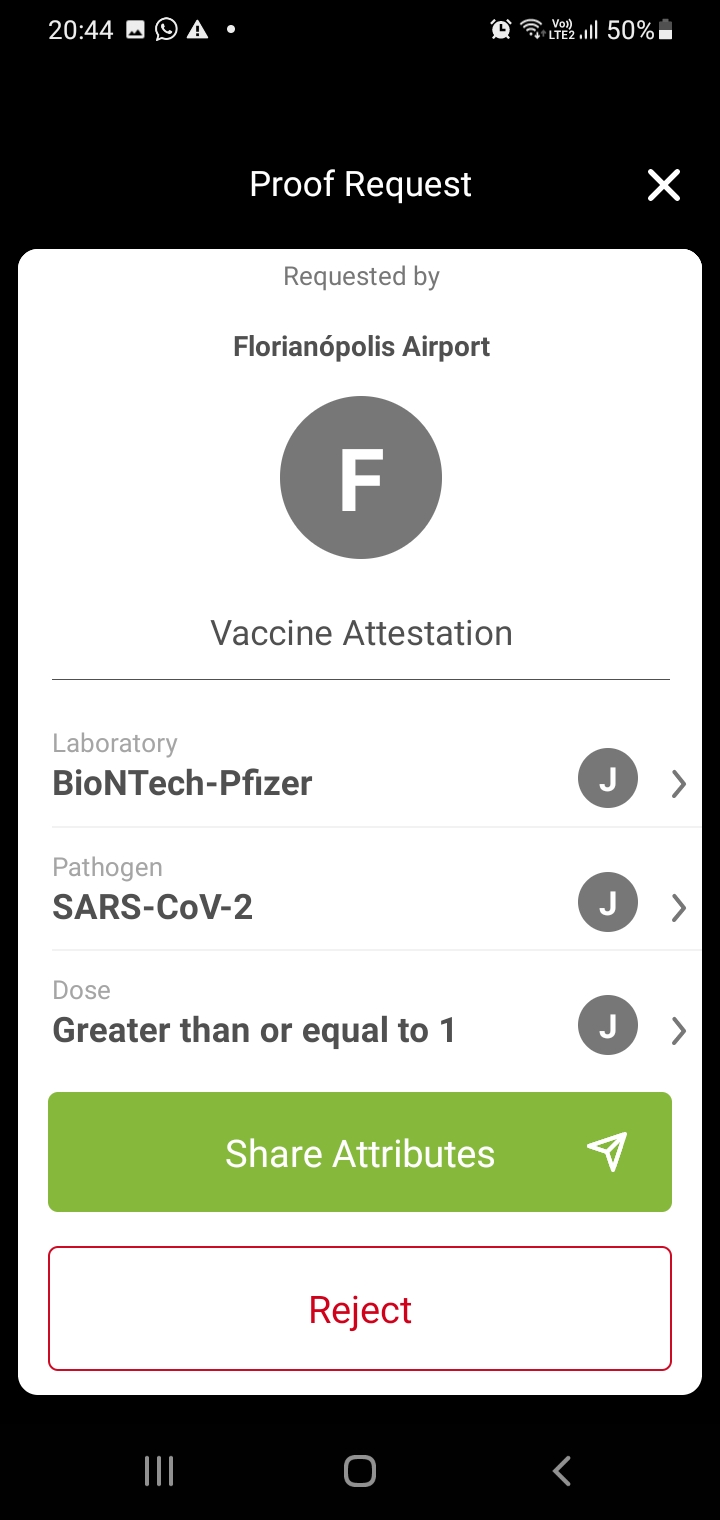}}
    \subfigure[Digital wallet is unable to create the requested VP.]{\label{fig:d}\includegraphics[width=41mm]{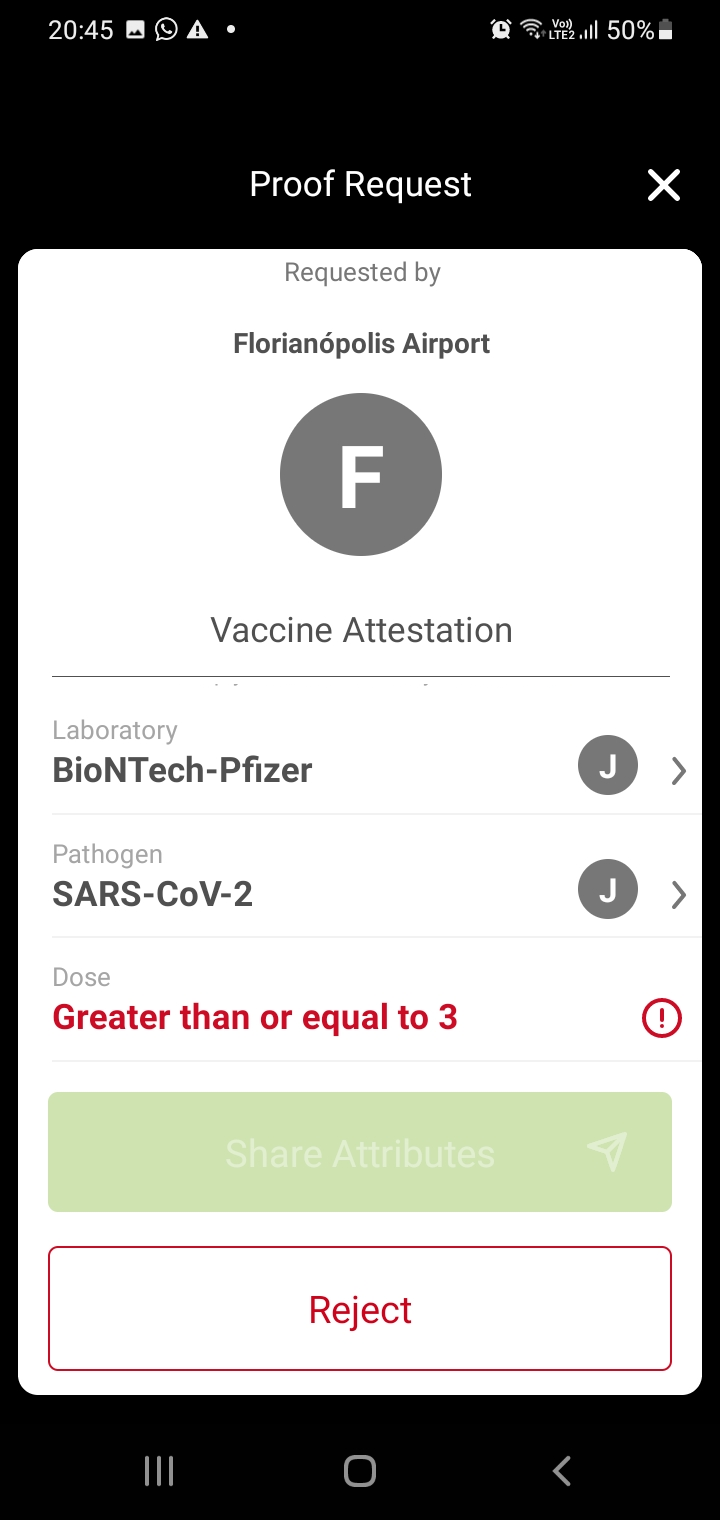}}
    \caption{Interactions between vaccinee and verifier through the digital wallet.}
\end{figure}

The interaction between vaccinee and verifier also begins through QR code, which we omit for lack of space. Figure~\ref{requisicaoProva} shows the verifier's interface, which defines a VP request format. The vaccinee will need to present the laboratory and pathogen their vaccinee fights, and the dose number must be greater than or equal to 1. Figure~\ref{fig:c} shows how the digital wallet shows the VP request to the vaccinee.

Finally, we show in Figure~\ref{fig:d} a VP request that the vaccinee cannot produce. In this case, the vaccinee is asked to prove that she has taken three or more vaccine doses for SARS-CoV-2. However, since no VC satisfies this requirement, the digital wallet cannot produce the VP for this request.

Selective disclosure and ZKP allow the holder to produce specific VPs for each context. With ZKP, it is possible to issue a credential and create fully anonymous VPs. Such anonymity makes no sense in a context where the holder is already identified, \textit{e.g.}, at work or university, and VPs can be adapted to that.

It is important to remark that all interactions of our system are contactless, helping with the necessary distancing measures to combat the pandemic.

\section{Related Work}\label{sec:related}

In early 2021, the Good Health Pass Collaborative (GHPC) was launched~\cite{ghpc2021} with more than 25 companies and organizations from multiple sectors. 
In the second half of 2021, the GHPC initiative released its blueprint~\cite{ghpcblue2021} with recommendations regarding credential and operational infrastructure. Design considerations and technical choices are based on five pillars: (i) individuals must be at the center of data exchange; (ii) the credential must allow the individual to provide evidence of their vaccination status; (iii) a decentralized approach is needed for global security and scalability; (iv) open standards are essential for interoperability and participation; and (v) pragmatic and realistic approach. The authors agree and adopt the recommendations of the GHPC for the elaboration of this work.

In a similar effort, the World Health Organization (WHO) created the Digital Documentation of COVID-19 Certificates (DDCC) intending to establish standards for an architecture for digital vaccination certificates~\cite{who2022}. In the second half of 2021, the DDCC published a report~\cite{whodocument2021} with the technical specifications and implementation guide for the COVID-19 certificates. The document assumes an existing Public Key Infrastructure (PKI) for each member state and proposes a digital counterpart to the paper certificate. It recognizes the importance of a global trust framework but does not detail how to implement a global health trust framework to store the public keys of member states. It is outside the report's scope to present technical features for the prospect of selective disclosure. So, unlike our approach, DDCC is not concerned with implementing a global chain of trust or a certificate that cares about data anonymity.

Similar to the WHO initiative, the European Union created the EU Digital COVID Certificate~\cite{eupass2021}. This initiative also uses PKI and certificates whose data come in plain text, without anonymity. The approach is feasible within the European Union with its own PKI. Still, acceptance in foreign countries is uncertain, depending on case-by-case interoperability agreements between nations.

In addition to national and international institutions, several private companies have also created their versions of health passes, such as the BLOK Pass~\cite{blockpass2021} and the Evernym Travel Pass~\cite{travelpass2021}. Both initiatives follow SSI principles and keep users' data only on their smartphones. However, a digital vaccination pass must be open-source, enabling code transparency and making it possible for anyone to verify the system.

Regarding academic efforts with similar objectives to this work, we present three research papers and point out their shortcomings.

The authors of~\cite{9105054} acknowledge that proof of vaccination or robust antibody testing will be in high demand and question what format such a certificate should take. They claim that a digital certificate would make more sense as long as it is: (i) privacy-preserving; (ii) un-forgeable; (iii) easy to administer; (iv) easy to verify while preserving privacy; (v) scalable to millions of users; and (vi) cost-effective. To achieve these goals, the authors propose a mobile application that implements an architecture based on: (i) VC; (ii) the Solid~\cite{solid2021} decentralized data platform, which stores the VCs issued on the user's smartphone; and (iii) a private Ethereum blockchain that uses proof of authority~\cite{curran2018proof}, where nodes are previously registered and authorized to confirm transactions. This solution, however, needs to register each user and each issued certificate in the blockchain, resulting in a high amount of transactions. Our solution does not require the blockchain for vaccinee registration and VC issuance, as it uses direct communication between the vaccinator and the vaccinee via DIDComm.

In~\cite{9286584} the authors seek to solve the challenge of proposing an immunity pass using Blockchain, SSI, and re-encryption proxies~\cite{Ateniese2006}, which allow an encrypted message with the public key of $A$ to be decrypted with the private key of $B$ through a proxy that re-encrypts the ciphertext. The solution features smart contracts for the Ethereum Blockchain~\cite{ethereum2021}. The patient's smart contract stores the hash of the vaccination and immunity records and travel history to perform contact tracing. However, as it is based on smart contracts in the Ethereum network, transactions have costs, although small, making it necessary for those involved to have financial resources available for execution.

Lastly, the authors of~\cite{articleCovid} chose to demonstrate the use of blockchain to create immunity certificates for SARS-CoV-2 in a pre-vaccine period with a document that certifies that the individual has been infected and is immune. The authors claim that the use of ``immunity licenses'' could create restrictions on who can and cannot interact in social spaces, encouraging counterfeiting. Therefore, the paper proposes using a government-operated blockchain registry, although it does not specify which one. After an antibody test, the details are kept in a smart contract confidentially. Their proposal uses biometric authentication as a private key for the ``account'' of patients on the blockchain, and their data is encrypted with the biometric data of the tested patient. However, the work does not offer any empirical evidence of the functioning of the proposed system.

\section{Final Remarks}\label{sec:conclusion}

SSI brings greater privacy, security, and ownership to user data than previous identity models. This work presents a system architecture that allows the issuance and verification of VCs based on SSI for proof of vaccination. Our implementation produces a VC with vaccination information that, through selective disclosure and ZKP, ensures proof of vaccination with a high level of privacy. While no single solution will be universally appropriate, researchers and practitioners can easily customize our system for different use cases by exchanging components such as digital wallets, blockchain, and software agents to their needs.


\printcredits

\bibliographystyle{cas-model2-names}

\bibliography{cas-refs}

\end{document}